\documentclass[reprint,amsmath,amssymb,aps]{revtex4-2}

\usepackage[T1]{fontenc}
\usepackage[utf8]{inputenc}
\usepackage{graphicx}
\usepackage{dcolumn}
\usepackage{bm}
\usepackage{amsfonts}
\usepackage{hyperref}

\begin{document}

\title{Spectroscopy of Heat Transport and Violation of the Wiedemann--Franz Law in a GaAs Hydrodynamic Mesoscopic Channel}

\author{Yu.\ A.\ Pusep}
\email{pusep@ifsc.usp.br}
\affiliation{S\~ao Carlos Institute of Physics, University of S\~ao Paulo, IFSC-USP, 13566-590 S\~ao Carlos, SP, Brazil}

\author{M.\ A.\ T.\ Patricio}
\affiliation{Physics Department, Federal University of S\~ao Carlos, 13565-905, S\~ao Carlos, SP, Brazil}

\author{M.\ M.\ Glazov}
\affiliation{Ioffe Institute, 194021 St.\ Petersburg, Russia}

\author{V.\ A.\ Oliveira}
\affiliation{Physics Department, Federal University of S\~ao Carlos, 13565-905, S\~ao Carlos, SP, Brazil}

\author{M.\ D.\ Teodoro}
\affiliation{Physics Department, Federal University of S\~ao Carlos, 13565-905, S\~ao Carlos, SP, Brazil}

\author{A.\ D.\ Levin}
\affiliation{Institute of Physics, University of S\~ao Paulo, 135960-170 S\~ao Paulo, SP, Brazil}

\author{A.\ K.\ Bakarov}
\affiliation{Institute of Semiconductor Physics, 630090 Novosibirsk, Russia}

\author{G.\ M.\ Gusev}
\affiliation{Institute of Physics, University of S\~ao Paulo, 135960-170 S\~ao Paulo, SP, Brazil}

\date{March 3, 2026}

\begin{abstract}
The Wiedemann--Franz law, which determines the universality of the ratio of
thermal conductivity to electrical conductivity, is studied in the
hydrodynamic electron transport regime, where electron--electron scattering
predominates over scattering by disorder. In this case, the different
relaxation of electric and thermal currents can lead to a violation of the
Wiedemann--Franz law, which is expected to be even more pronounced in
mesoscopic electron systems. This paper reports the propagation of hot
electrons in a GaAs hydrodynamic narrow channel, studied using
micrometer-resolution photoluminescence thermometry. A temperature dependence
of the Lorenz number was obtained, indicating a violation of the
Wiedemann--Franz law. The important role of narrow constrictions in this
violation was also demonstrated, and theoretical arguments are presented.
\end{abstract}

\maketitle

\section{Introduction}

Over the past decade, considerable attention has been devoted to the thermal
properties of clean correlated electron systems in which momentum-conserving
electron--electron (e--e) scattering dominates over momentum-relaxing
scattering by disorder. In this regime, the single-particle approximation is
no longer adequate, and the collective dynamics of electrons must be taken
into account. This behavior is naturally described within the hydrodynamic
approach. One of the most intriguing phenomena expected in hydrodynamic
electron systems is the violation of the Wiedemann--Franz (WF) law, a relation
that is otherwise remarkably robust in solid-state physics. A pronounced
violation of the WF law therefore strongly suggests unconventional physics and,
in electronic systems, may indicate hydrodynamic transport
\cite{narozhny2022,fritz2024,hui2025}.

The WF law states that, in metals at temperatures $T$ much lower than the
Fermi energy, the ratio of electronic thermal conductivity $\kappa_e$ to
electrical conductivity $\sigma$ can be written in universal form through the
Lorenz number:
\begin{equation}
L_{0}=\frac{\kappa_{e}}{\sigma T}
=\frac{\pi^{2}}{3}\left(\frac{k_{B}}{e}\right)^{2}.
\label{eq:L0}
\end{equation}
Here $e$ is the elementary charge and $k_{B}$ is the Boltzmann constant.
Equation~\eqref{eq:L0} gives the Sommerfeld value, originally derived within
the classical Drude theory of electron transport \cite{eckert1987}. It
reflects the fact that the same carriers, electrons, are responsible for both
charge and heat transport. Moreover, within Fermi-liquid theory, the Lorenz
number remains universal provided that the system stays metallic
\cite{castellani1987}. Thus, the WF law is expected to hold as long as
Fermi-liquid theory applies and inelastic scattering does not distinguish
between charge and heat currents. Since inelastic scattering generally
vanishes as $T\to0$, the WF law is expected to apply to metallic electron
systems in the zero-temperature limit.
\begin{figure*}[t]
    \centering
    \includegraphics[width=12 cm]{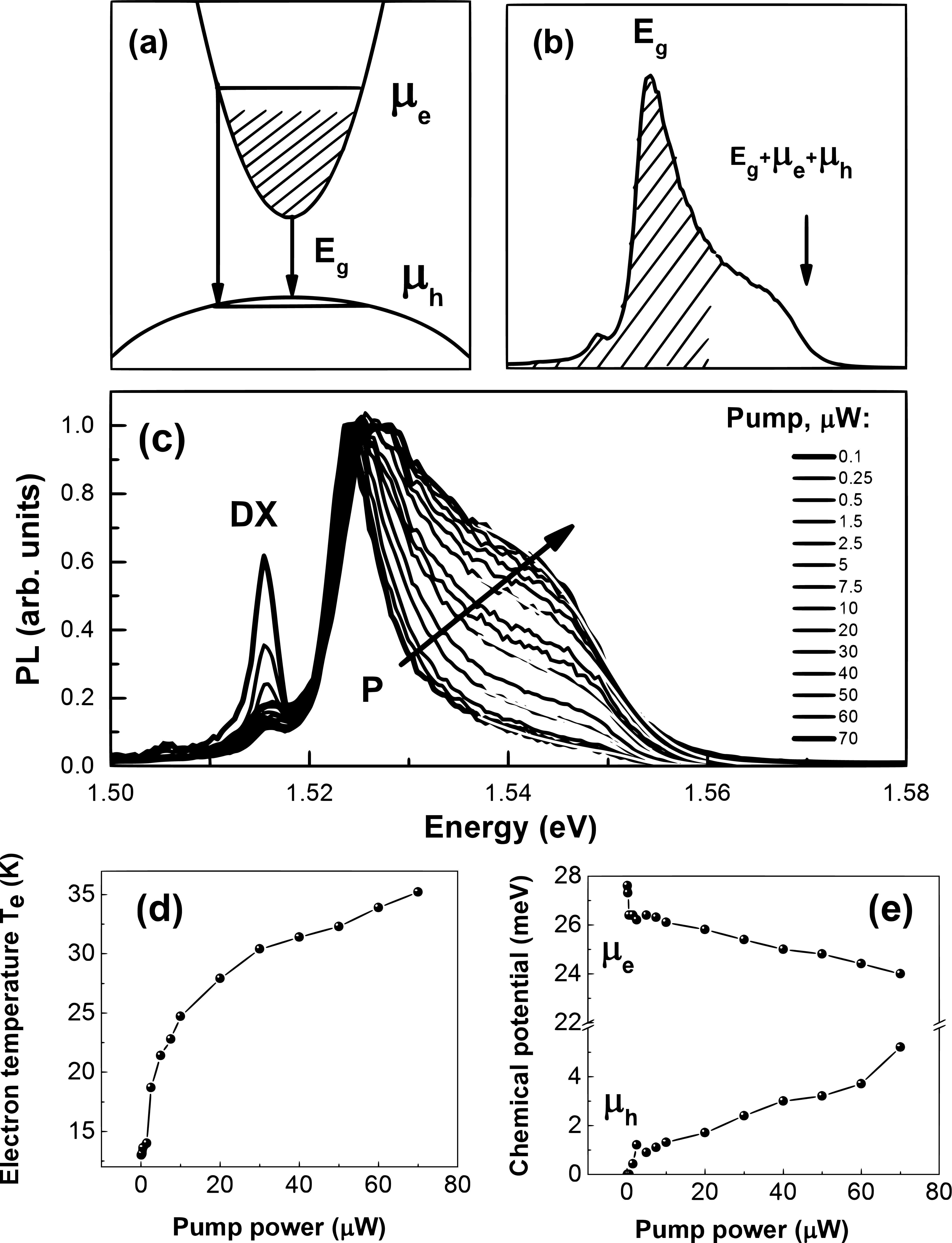}
    \caption{\textbf{Principle of the PL thermometry.}
    (a) Interband transitions.
    (b) Resulting PL spectrum.
    The shaded areas represent temperature-insensitive states in the conduction band in (a)
    and their contribution to PL in (b), where the high-energy tail of the PL, which is
    affected by temperature, is shown in red.
    (c) PL spectra measured in the GaAs channel at different pump powers $P$ at $4~\mathrm{K}$,
    where the cyan lines are best fits using Eq.~(3).
    Electron temperature (d) and the chemical potentials of electrons and holes (e).}
    \label{fig:fig1}
\end{figure*}
The situation changes qualitatively when, in addition to renormalization of
the energy spectrum, e--e collisions are explicitly taken into account.
Although e--e collisions conserve the total momentum of the interacting pair,
they do not conserve the energy of each individual electron. As a result,
e--e scattering suppresses thermal conductivity and can therefore lead to
deviations from the universal WF law, Eq.~\eqref{eq:L0}
\cite{pomeranchuk1950,abrikosov1957,baym1991}. The key reason is that charge
and heat currents are governed by different components of the nonequilibrium
distribution function. The charge current is directly proportional to the
total momentum of the electron system. Because e--e collisions conserve total
momentum, they do not relax the charge current. By contrast, the heat current
depends on energy flow and involves higher angular harmonics of the
distribution function. During e--e collisions, energy is redistributed among
electrons, and the directional components of the energy flux are partially
randomized. As a consequence, the second and higher angular harmonics of the
distribution function relax efficiently. Although the total energy of the
system is conserved, directed heat transport decays because of this angular
redistribution. Therefore, when e--e collisions dominate over
momentum-relaxation processes, that is, in the hydrodynamic regime, inelastic
scattering strongly affects the thermal current while leaving the electrical
current nearly unchanged. A suppressed Lorenz number is therefore expected,
and this has been identified as an intrinsic feature of hydrodynamic electron
systems \cite{principi2015,dassarma2018,dassarma2022,lee2020}.

Violation of the WF law in hydrodynamic two-dimensional (2D) electron systems
was first observed in graphene \cite{fong2013,crossno2016,aniket2025}, where
the relativistic Coulomb-interacting electron--hole plasma creates special
conditions. Near charge neutrality, at the Dirac point, charge transport
becomes ambipolar. In this regime, both electrons and holes contribute to the
electrical current. Near the charge-neutrality point, large Lorenz numbers
were reported, whereas in the degenerate Fermi-liquid regime Lorenz numbers up
to three orders of magnitude smaller than the Sommerfeld value were observed.
In graphene, electrons and holes are driven in opposite directions, and
electron--hole collisions limit the electrical current. By contrast, a
temperature gradient drives both electrons and holes in the same direction, so
the corresponding energy flux can grow without suppression by interparticle
collisions \cite{aleiner2009}. This leads to a thermal conductivity much
larger than that predicted by the WF law. Other possible origins of large
Lorenz numbers in graphene have also been discussed, including the linear
energy spectrum and quantum conductivity \cite{narozhny2022}. In addition, it
has been proposed that an apparent enhancement of the Lorenz number may arise
from the opening of a gap at the Dirac point rather than from a genuine
violation of the WF law \cite{dassarma2023}.
\begin{figure*}[t]
    \centering
    \includegraphics[width=18cm]{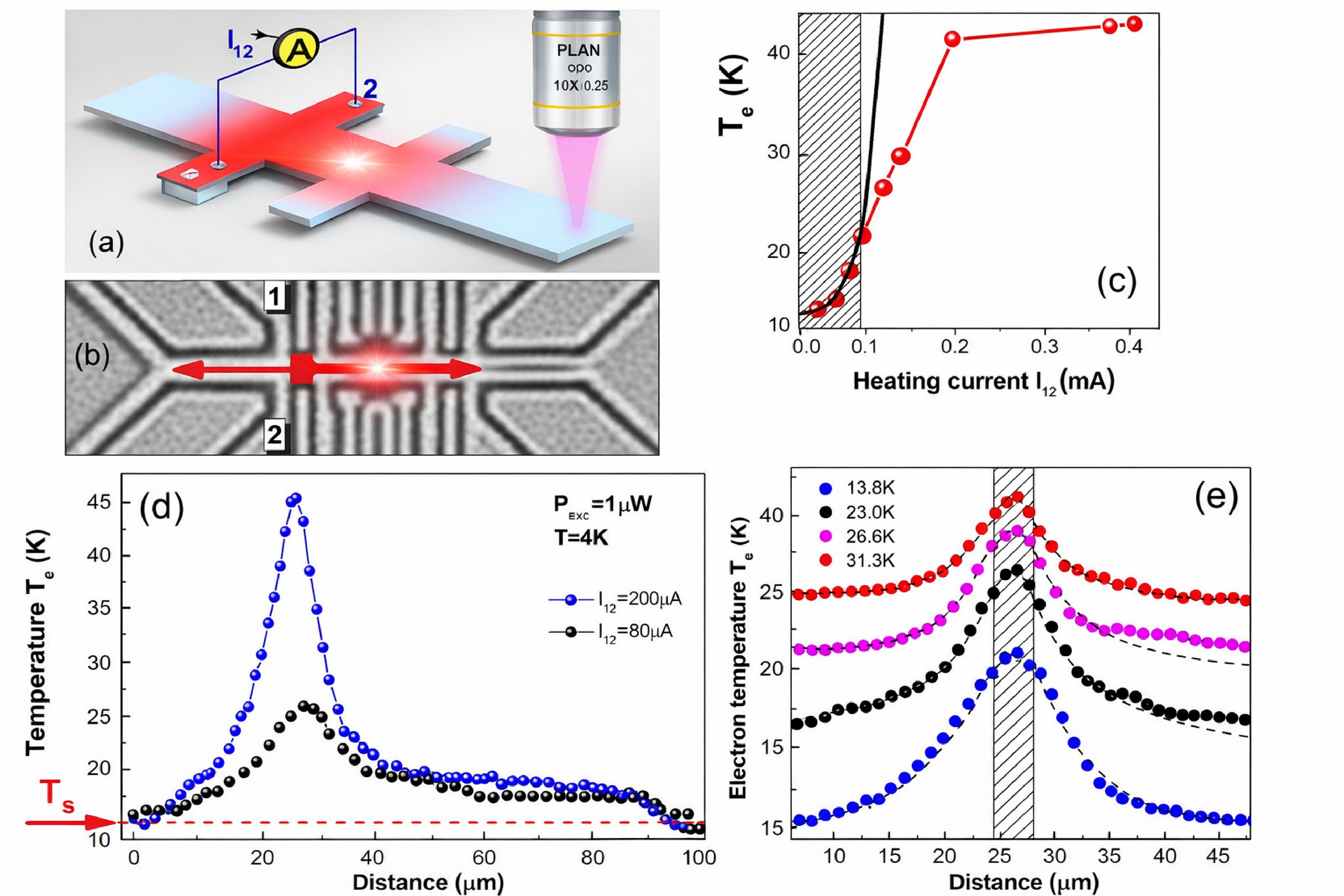}
    \caption{\textbf{Temperature of hot electrons.}
    (a) Measurement scheme.
    (b) Microscopic image of the sample where the heat-source area and directions of heat
    transfer are indicated by the red square and red arrows, respectively.
    (c) Electron temperature measured as a function of the heating current.
    (d) Temperature profiles measured along the channel at $T=4~\mathrm{K}$, at laser pump
    power of $1~\mu\mathrm{W}$ and heating currents of 80 and $200~\mu\mathrm{A}$;
    $T_s$ is the sample temperature.
    (e) Electron temperature profiles measured along the channel at a laser pump power of
    $2.5~\mu\mathrm{W}$ at different sample temperatures; the dashed lines represent best fits,
    as explained in the text, and the shaded area indicates the range between heating contacts
    where the heat power is assumed to be constant.}
    \label{fig:fig2}
\end{figure*}
In the experiments mentioned above, the electron temperature was measured
using sensitive Johnson-noise thermometry, where hot-carrier-induced noise is
associated with the electron temperature. However, in addition to thermal
noise, other contributions may be present, such as convective noise caused by
carrier-drift fluctuations, intervalley noise in multivalley semiconductors,
surface trapping, and generation--recombination noise \cite{bauer1974}. These
effects complicate the interpretation of the measurements. Moreover, in the
experimental geometries commonly used, charge and heat flow in parallel,
which can lead to collateral effects such as convection, the Seebeck effect,
and heat dissipation at electrical contacts. Another approach was used to
study heat transport in a graphene channel by measuring the amplitude of
universal conductance fluctuations using lateral probes \cite{draelos2019}.
However, that method only probes lateral heat transport along the channel
edges.

To the best of our knowledge, no direct evidence has yet been reported for the
validity or violation of the WF law in GaAs-based hydrodynamic electron
systems. In 2D GaAs, hydrodynamic effects are expected to be even more clearly
manifested than in graphene, and a violation of the WF law is predicted in
the temperature range $T \simeq 4$--$40~\mathrm{K}$ for samples with mobility
exceeding $10^{6}\,\mathrm{cm^{2}\,V^{-1}\,s^{-1}}$ \cite{dassarma2022}.
Thus, 2D GaAs is an excellent platform for investigating the violation of the
WF law. Moreover, the WF law has not yet been explored in hydrodynamic
mesoscopic electron systems satisfying $l_{ee}<w<l_{p}$, where $l_{ee}$,
$l_{p}$, and $w$ denote the e--e interaction length, the mean free path, and
the characteristic device size, respectively. In such systems, boundaries may
affect thermal and electrical conductivity in different ways, suggesting that
the violation of the WF law may be even stronger in the mesoscopic regime.

Here we report measurements of thermal transport in a mesoscopic channel based
on a high-mobility GaAs quantum well (QW) hosting a 2D electron gas. Our
approach combines direct optical measurements of the local electron
temperature with local Joule heating induced by a transverse current.
Together with transport measurements, this enables direct extraction of the
Lorenz number from fits to the temperature profile in a geometry without
parallel electrical current. We observe strong deviations of the measured
Lorenz number from the value predicted by Eq.~\eqref{eq:L0} and propose a
simple model incorporating both e--e scattering and boundary effects to
describe the observed violation of the WF law.

\section{Temperature of Hot Electrons in a Mesoscopic GaAs Channel}

In this paper, we report measurements of hot-electron propagation within a
hydrodynamic mesoscopic channel fabricated on the basis of a GaAs QW. Under
the action of an electric current flowing perpendicular to the channel through
potentiometric contacts, hot electrons are created in the channel cross
section between them. Their temperature is measured along the channel as a
function of distance from the heating source using a photoluminescence (PL)
setup with micrometric resolution. The electron temperature is measured by
fitting the high-energy PL tail, whose spectral shape is determined by the
distributions of electrons and holes in the conduction and valence bands,
respectively \cite{lyo1988,vezin2024}. Thus, the temperature profile of hot
electrons along the channel was obtained.

When a local heat power density $p$ is applied to an electron system,
resulting in a temperature gradient, the heat balance is given by the
one-dimensional heat-transfer equation
\begin{equation}
p = -\frac{d}{dx}\left(\sigma L T_{e}\frac{dT_{e}}{dx}\right)
+ \Sigma_{e\text{-}ph}\left(T_{e}^{\delta}-T_{ph}^{\delta}\right).
\label{eq:heat}
\end{equation}
Here $L$ and $\sigma$ are the Lorenz number and electrical conductivity,
respectively, whose temperature dependence can be neglected for weak heating
of the electron gas, and $\Sigma_{e\text{-}ph}$ is the electron--phonon
coupling constant. The first term describes thermal conductivity, that is,
energy transfer due to electron diffusion, while the second term describes
cooling due to phonon emission. The exponent $\delta \gtrsim 3$ typically
depends on the interaction mechanism \cite{ridley1991}.
\begin{figure}[t]
    \centering
    \includegraphics[width=\columnwidth]{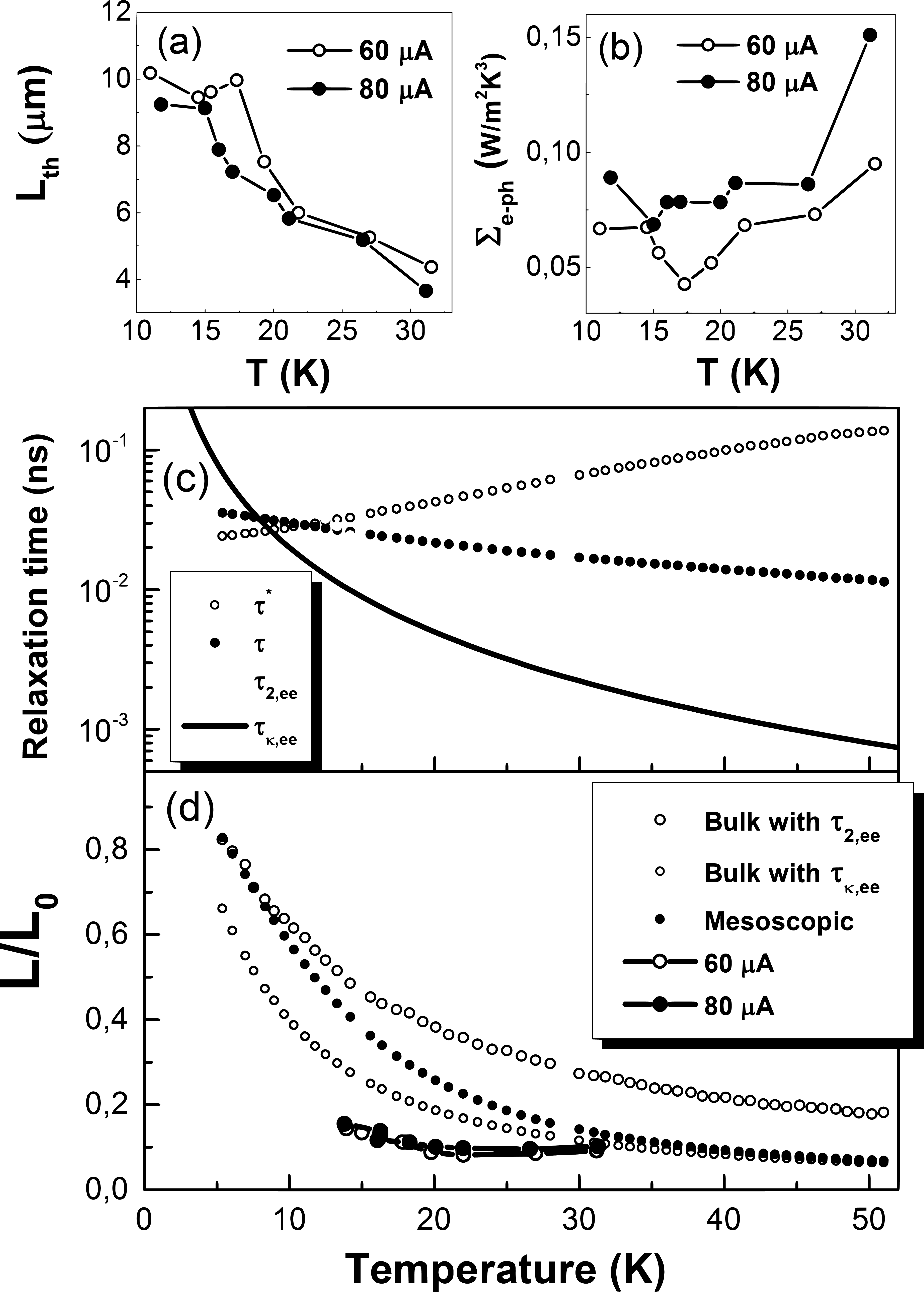}
    \caption{\textbf{Characteristic parameters of hydrodynamic electrons.}
    (a) Effective thermal length $L_{\mathrm{th}}$ and
    (b) electron--phonon coupling constant $\Sigma_{e\text{-}ph}$ as functions of temperature.
    (c) Characteristic relaxation times.
    (d) Lorenz ratio calculated using Eq.~(4) (black and blue open circles), as explained in
    the text, and Eq.~(6) (solid blue circles), and obtained by fitting the electron temperature
    profiles measured at heating currents of $60~\mu\mathrm{A}$ (red open circles) and
    $80~\mu\mathrm{A}$ (red solid circles) as a function of temperature.}
    \label{fig:fig3}
\end{figure}

\section{WF Law}

The Lorenz ratio was calculated according to
\cite{principi2015,dassarma2018}
\begin{equation}
\frac{L}{L_{0}}=\frac{1}{1+\tau/\tau_{ee}}.
\tag{4}
\end{equation}
Here $\tau$ and $\tau_{ee}$ are the momentum-relaxation time and the e--e
scattering time, respectively. The Lorenz ratio in Eq.~(4) reflects the fact
that e--e scattering promotes relaxation of the heat current but not the
charge current.

\section{Lorenz Ratio in a Mesoscopic Device}

Based on previous works \cite{alekseev2016,gurzhi1963,gurzhi1968}, we derive
the following expression for the electrical conductivity:
\begin{equation}
\sigma=\frac{n e^{2}}{m}\frac{\tau\tau^{\ast}}{\tau+\tau^{\ast}}.
\tag{5a}
\end{equation}
This expression takes into account both electron momentum scattering $\tau$,
caused by impurities and phonons, and the additional resistance channel
related to electron viscosity $\eta$, described by the time
$\tau^{\ast}=w^{2}/(12\eta)$, where $w$ is the channel width
\cite{alekseev2016}.

The electron thermal conductivity can be expressed in a similar form:
\begin{equation}
\kappa_{e}
=\frac{\pi^{2}}{3}T\frac{n k_{B}^{2}}{m}
\frac{\widetilde{\tau}\tau^{\ast}}{\widetilde{\tau}+\tau^{\ast}}.
\tag{5b}
\end{equation}
Here $\widetilde{\tau}$ includes the heat-current relaxation caused by e--e
collisions,
\begin{equation}
\widetilde{\tau}=\frac{\tau\tau_{\kappa,ee}}{\tau+\tau_{\kappa,ee}},
\end{equation}
where $\tau_{\kappa,ee}$ is the relaxation time associated with e--e
collisions.

Thus, the mesoscopic Lorenz ratio is
\begin{equation}
\frac{L}{L_{0}}
=\frac{\widetilde{\tau}}{\tau}
\frac{\tau+\tau^{\ast}}{\widetilde{\tau}+\tau^{\ast}}.
\tag{6}
\end{equation}

\section{Discussion and Conclusions}

The ``bulk'' Lorenz ratio obtained from Eq.~(4), when evaluated using the
e--e relaxation time $\tau_{\kappa,ee}$, is in better agreement with the
experimental data. This occurs because, in the relevant temperature range,
$\tau^{\ast}>\tau$, and therefore the Poiseuille flow is not yet fully
developed in the channel. Even in this regime, however, the existence of two
distinct relaxation times, $\tau$ and $\tau_{\kappa,ee}$, leads to clear and
measurable deviations from the standard WF law.

We emphasize that the experimental Lorenz ratio extracted from the measured
electron-temperature profiles differs from the Lorenz ratio calculated from
Eqs.~(4) and (6) using relaxation times obtained from magnetoresistance
measurements. In the former case, the electron temperature is determined in a
hot-electron regime, whereas in the latter case the relaxation times are
measured with electrons in thermal equilibrium with the lattice. Therefore,
the Lorenz number extracted from the temperature profiles should be regarded
as an average value for an ensemble of hot electrons. A reasonable estimate of
the average hot-electron temperature is given by the temperature at half of
the maximum of the temperature profile. On this basis, the experimental
Lorenz-ratio data should be shifted by approximately $4$--$7~\mathrm{K}$
toward higher temperatures for comparison with the calculated curves. With
this correction, the agreement between theory and experiment is satisfactory.

In summary, electron heat transport was investigated by combining optical
spectroscopy and electrical transport measurements. The electron-temperature
profile as a function of distance from a local heat source was measured in a
mesoscopic GaAs channel containing a high-mobility 2D electron gas operating
in the hydrodynamic regime, using micrometer-resolution photoluminescence
thermometry. This approach allowed us to extract the Lorenz ratio directly and
to demonstrate a pronounced violation of the Wiedemann--Franz law. The results
can be explained within a model that incorporates the role of e--e scattering
in both charge and heat transport and implies different e--e scattering times
for the relaxation of electrical and thermal currents.

\begin{acknowledgments}
Financial support from the Brazilian agency FAPESP (Grants 2021/12470-8 and
2022/10340-2) is gratefully acknowledged. M.\ M.\ G.\ is grateful to RSF Grant
No.~22-12-00211-Continuation (analytical model).
\end{acknowledgments}


\clearpage
\onecolumngrid

\setcounter{section}{0}
\setcounter{subsection}{0}
\setcounter{equation}{0}
\setcounter{figure}{0}
\setcounter{table}{0}

\renewcommand{\thesection}{S\arabic{section}}
\renewcommand{\thesubsection}{S\arabic{section}.\arabic{subsection}}
\renewcommand{\theequation}{S\arabic{equation}}
\renewcommand{\thefigure}{S\arabic{figure}}
\renewcommand{\thetable}{S\arabic{table}}

\section*{Supplemental Material}
\addcontentsline{toc}{section}{Supplemental Material}

This supplemental material provides detailed information on the sample
characteristics and the measurement methods used in the main text.

\section{Hydrodynamic Transport Regime}

The hydrodynamic approach to electron behavior in two-dimensional fermionic
systems offers a unique perspective that diverges from traditional kinetic
theory, revealing fascinating predictions for electron transport,
particularly in small-scale samples. A key insight is that, when
electron--electron interactions are strong enough, the system can be described
by a viscous hydrodynamic framework, allowing for new interpretations of
transport phenomena. Recent breakthroughs in materials science, especially in
producing exceptionally clean samples, have enabled researchers to
systematically explore these hydrodynamic effects across various
two-dimensional electronic systems. Hydrodynamic electron flows are
anticipated in transport phenomena when the mean free path for
electron--electron collisions (denoted as $l_{ee}$) is significantly shorter
than the mean free path due to impurity and phonon scattering (represented as
$l$).

\subsection{Measurement Details}

We fabricated our devices using high-quality GaAs quantum wells, each with a
width of 14 nm and an electron density of approximately
$9.1\times10^{11}\,\mathrm{cm^{-2}}$ at 4.2 K. The macroscopic sample
demonstrated a mobility of $2\times10^{6}\,\mathrm{cm^{2}\,V^{-1}\,s^{-1}}$.
For our measurements, we designed a Hall bar specifically suited for
multiterminal experiments, comprising three consecutive segments with lengths
of 6, 20, and $6\,\mu\mathrm{m}$, all with a width of $6\,\mu\mathrm{m}$.
We also integrated eight voltage probes into this setup. Ohmic contacts to the
two-dimensional electron system were created by annealing Ti/Ni/Au layers
deposited on the GaAs surface. For our measurements, we used a VTI cryostat
combined with a standard lock-in detection technique to measure longitudinal
resistance. To prevent overheating, we applied an alternating current (AC) in
the range of $0.1$--$1\,\mu\mathrm{A}$, a level considered sufficiently low
for these tests.

\begin{figure}[ht]
\centering
\includegraphics[width=0.8\linewidth]{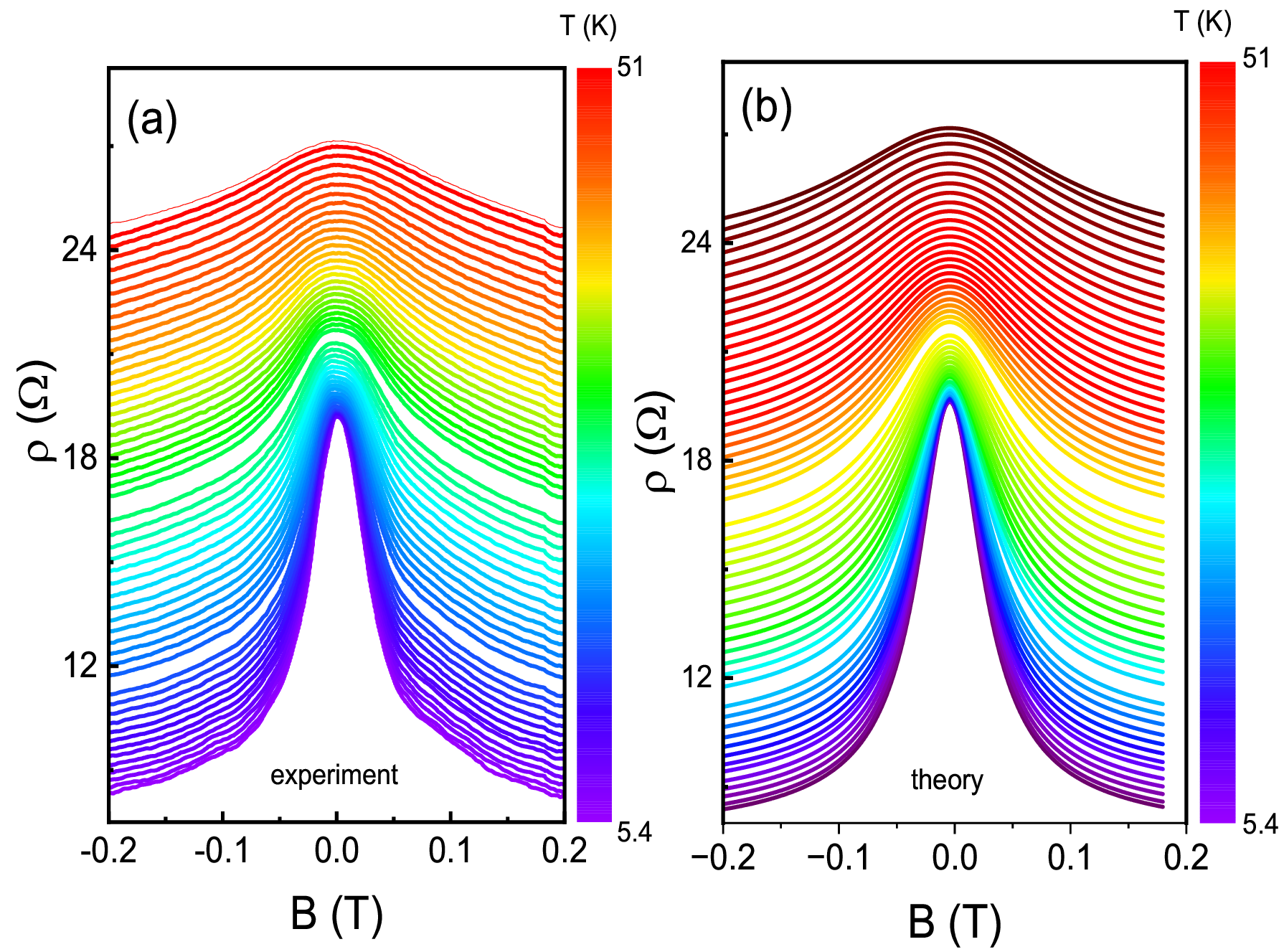}
\caption{Temperature-dependent magnetoresistivity: (a) experiment and
(b) theoretical curves calculated from Eq.~(\ref{eq:S2}) for different
temperatures.}
\label{fig:S1}
\end{figure}

Figure~\ref{fig:S1} illustrates the variation in resistivity,
$\rho=(W/L)R$, with magnetic field at different temperatures. A notable
characteristic in these samples is the pronounced negative magnetoresistivity,
$\rho(B)-\rho(0)<0$, which follows a Lorentzian profile. As temperature
increases, this negative magnetoresistivity diminishes in magnitude and
broadens. In addition, the resistivity at zero magnetic field rises with
temperature.

\subsection{Theory and Discussion}

To qualitatively compare with the experimental data from samples without
obstacles, we apply a model from previous research, initially designed to
describe Poiseuille flow under the influence of a magnetic field
\cite{alekseev1,scaffidi}.

\begin{figure}[ht]
\centering
\includegraphics[width=0.8\linewidth]{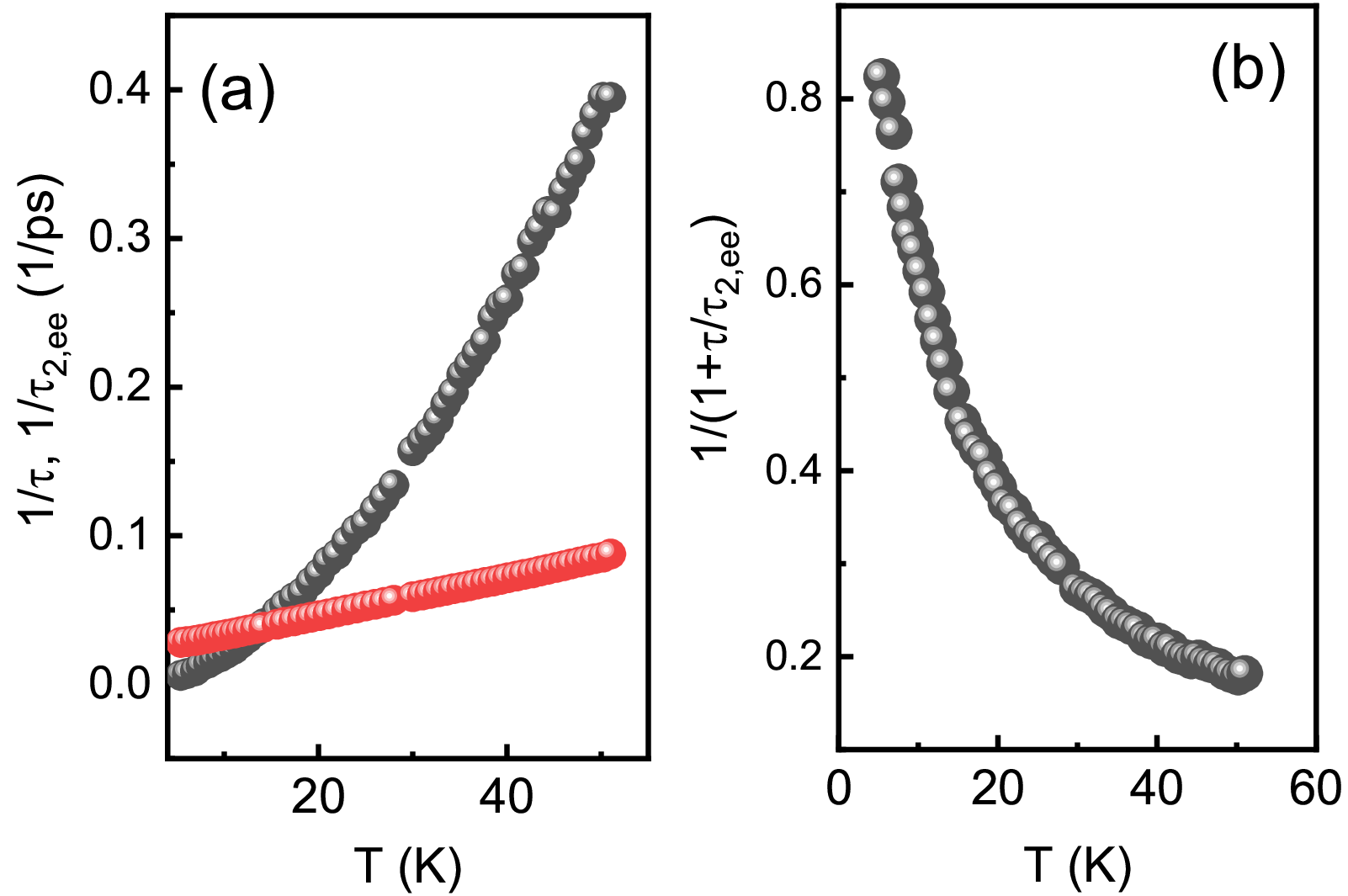}
\caption{(a) The relaxation rates $1/\tau$ (red circles) and
$1/\tau_{2,ee}$ (black circles) as functions of temperature.
(b) The ratio $1/(1+\tau/\tau_{2,ee})$ as a function of temperature.}
\label{fig:S2}
\end{figure}

In a simplified form, the model describes resistivity as the result of two
main contributions. The first stems from ballistic effects or scattering due
to boundaries and defects, while the second is governed by viscosity
\cite{alekseev1}:
\begin{equation}
\rho(B)= \frac{m}{e^{2}n}\left(\frac{1}{\tau}+\frac{1}{\tau^{*}}\right).
\label{eq:S1}
\end{equation}
Here, $1/\tau$ represents the scattering rate due to static disorder, while
$m=0.067\,m_{0}$ and $n$ denote the effective mass and density, respectively,
with $m_0$ being the electron mass. The relaxation time is
$\tau^{*}=W^{*2}/(12\eta)$, where $\eta=\frac{1}{4}v_{F}^{2}\tau_{2}$ is the
viscosity. The term $W^{*}$ refers to the effective sample width, which, in
the case of a no-slip boundary condition, matches the geometric width $W$.
The relaxation rate $\tau_{2}$ corresponds to the shear-stress relaxation time
arising from electron--electron scattering. The subscript ``2'' signifies that
the viscosity coefficient is governed by the relaxation of the second harmonic
in the distribution function \cite{alekseev1}.

For a more complete formulation in a magnetic field, the theory incorporates a
viscosity tensor, which is dependent on the magnetic field, to determine the
resistivity tensor:
\begin{equation}
\rho(B)=\left(\frac{m}{e^{2}n\tau}\right)\frac{1}{1-\tanh(\xi)/\xi}.
\label{eq:S2}
\end{equation}
In this context, the dimensionless Gurzhi parameter is defined as
$\xi=\xi_{0}\sqrt{1+(2l_{2}/r_{c})^{2}}$, where $\xi_{0}=W/l_{G}$, with
$l_{G}=\sqrt{l_{2}l}$ representing the Gurzhi length. Here,
$l_{2}=v_{F}\tau_{2}$, $l=v_{F}\tau$, and $r_{c}=v_{F}/\omega_{c}$ is the
cyclotron radius. The cyclotron frequency is $\omega_{c}=eB/(mc)$. The
shear-viscosity relaxation rate is given by
\begin{equation}
\frac{1}{\tau_{2}(T)}=\frac{1}{\tau_{2,ee}(T)}+\frac{1}{\tau_{2,\mathrm{imp}}},
\label{eq:S3}
\end{equation}
while the momentum-relaxation rate is expressed as
\begin{equation}
\frac{1}{\tau(T)}=\frac{1}{\tau_{0,\mathrm{ph}}(T)}+\frac{1}{\tau_{0,\mathrm{imp}}}.
\label{eq:S4}
\end{equation}
In this expression,
$1/\tau_{0,\mathrm{ph}}=B_{\mathrm{ph}}T$ corresponds to phonon scattering,
and $1/\tau_{0,\mathrm{imp}}$ represents scattering due to static disorder,
distinct from the relaxation time for the second harmonic \cite{alekseev1}.

We then fit the magnetoresistance curves and the resistivity $\rho(T)$ at zero
magnetic field. The fitting procedure employs three parameters: $\tau(T)$,
$\tau_{2}(T)$, and the sample width $W^{*}$. Let us examine the data
regarding electron--electron interactions and relaxation caused by static
disorder, as derived from magnetoresistance analysis. Figure~\ref{fig:S2}
illustrates the temperature-dependent behavior of the corresponding relaxation
rates. To facilitate comparison with theoretical predictions, we used the
parameters $1/\tau_{2,\mathrm{imp}}$, $1/\tau_{0,\mathrm{imp}}$, $A_{ee}$,
and $B_{ph}$, as listed in Table~\ref{tab:S1}. Employing
Eq.~(\ref{eq:S3}), the e--e relaxation rate is expressed as
\begin{equation}
\frac{\hbar}{\tau_{2,ee}} = A_{ee}\frac{(kT)^2}{E_F}.
\label{eq:S5}
\end{equation}

\begin{table}[ht]
\caption{Fitting parameters of the electron system. The parameters are defined
in the text.}
\label{tab:S1}
\begin{ruledtabular}
\begin{tabular}{lccccc}
& $1/\tau_{2,\mathrm{imp}}$ & $1/\tau_{0,\mathrm{imp}}$ & $A_{ee}$ & $B_{ph}$ & $W^{*}$ \\
& $(10^{11}\,\mathrm{s^{-1}})$ & $(10^{10}\,\mathrm{s^{-1}})$ & & $(10^{9}\,\mathrm{s^{-1}K^{-1}})$ & $\mu\mathrm{m}$ \\
\hline
& 1.7 & 2.8 & 0.6 & 1.3 & 8.4 \\
\end{tabular}
\end{ruledtabular}
\end{table}

It can be observed that all relaxation rates converge onto universal curves:
$1/\tau_{2,ee}\sim T^2$ and $1/\tau\sim T$. One can also see that the
effective width is larger than the geometric width. We attribute this
discrepancy to the finite slip length caused by specific scattering at the
boundaries. A theoretical model \cite{gromov} proposes that
$W^{*2}=W(W+6l_s)$, which indeed predicts a larger effective width for samples
with a finite slip length. By comparing with this model, we estimate
$l_s\sim1$--$2\,\mu\mathrm{m}$. For diffusive boundary scattering, the
velocity distribution profile in the channel is parabolic, corresponding to
Poiseuille flow in a liquid. The slip length is the distance where the
extrapolated velocity vanishes \cite{kiselev,raichev}. A finite slip length
modifies the velocity distribution, shaping it as a ``cut parabola''
\cite{gusev,kiselev}.

Based on these calculations, we can discuss the conditions for hydrodynamic
effects in our samples. The hydrodynamic description is applicable under
conditions where $1/\tau < W/v_F < 1/\tau_{2,ee}$, with $W$ the sample width.
This indicates that we remain within the hydrodynamic regime even at
$T > 10\,\mathrm{K}$. Experimentally, this is supported by
Fig.~\ref{fig:S2}. In addition, we calculated the ratio
$1/(1+\tau/\tau_{2,ee})$, which is an important factor for considering the
violation of the Wiedemann--Franz law, as mentioned in the main text. One can
see that this ratio is less than 1 and decreases with increasing temperature.

\end{document}